\documentstyle[pra,aps]{revtex}
\begin{document}
\draft

\title {Analytical solutions to one-dimensional
dissipative and discrete chaotic dynamics}

\author{D.~Pingel and P.~Schmelcher}
\address{Theoretische Chemie, Physikalisch-Chemisches Institut, Universit\"at
Heidelberg, Im Neuenheimer Feld 253\\
D-69120 Heidelberg, Germany}

\author{F.K.~Diakonos}
\address{Department of Physics, University of Athens, GR-15771 Athens, Greece}

\date{\today }
\maketitle

\begin{abstract}
Analytical solutions to the chaotic and ergodic motion of a certain class of
one-dimensional dissipative and discrete dynamical systems are derived.
This allows us to obtain exact expressions for physical properties like
the time correlation function. We illustrate our solutions by means of
a few examples for which conventional numerical trajectory calculations
fail to predict the correct behaviour. 
\end{abstract}


\section{Introduction}
Chaotic behaviour is a characteristic feature of the
overwhelming majority of deterministic nonlinear dynamical systems \cite{Schu94}.
Classical chaos manifests itself in the exponential sensitivity
of the trajectories with respect to the initial conditions
and has nicely been understood as the very complex topological process of 
ever lasting stretching and folding of the motion.
In view of the complexity of chaotic trajectories it appears hard to
imagine that chaotic motion could be described by means of closed analytical
formulae. To our knowledge there exists only one exception, i.e. dynamical
law, which allows the representation of its chaotic trajectories for 
arbitrarily long times in a closed analytical form: the solutions
of the logistic map in the ergodic chaotic limit are given by
the celebrated Pincherl$\acute{e}$ relation \cite{Pin20}.
In contrast to our lack of knowledge concerning exact expressions for
chaotic motion there exists a well-founded and justified interest
in obtaining solutions to ergodic and chaotic behaviour.

On the o.h.s. it is clear that the frequently used numerical solutions
of chaotic dynamical systems do not yield one and the same trajectory 
for long time scales \cite{Cau94}. Nevertheless due to the shadowing argument
statistical quantitites of chaotic ergodic systems with and without
external noise can in many cases approximately be obtained
through a numerical investigation \cite{Gre90,Sau97}.
However there are exceptions like the case when a Lyapunov exponent
fluctuates about zero and this is expected to be common in
simulations of higher-dimensional systems \cite{Daw94}.
It is therefore highly desirable to make exact properties accessible:
if the exact trajectory could be derived this would be an excellent
starting-point for calculating the exact correlation function, invariant
measure, Lyapunov exponent or other quantities \cite{Dia96}. In particular it would
also offer the possibility of determining the exact long time behaviour
of relevant physical quantities. Even more important may be the 
fact that analytical representations
of chaotic motion are of principle interest and can certainly
enhance our understanding of the complexity of chaotic dynamics
in general. Typical questions which could then be addressed are:
How does the exponential sensitivity with respect to the initial
conditions come about? Are there any 
characteristic (scaling) properties and/or self-similar structures
of the analytical expressions
which are responsible for the complex stretching and folding process
of chaotic dynamics?

The current investigation represents a first step in the above direction
and provides closed analytical formulae for two classes of one dimensional unimodal,
dissipative, ergodic and chaotic maps of the interval each containing an infinite number of members.
In particular we will derive exact solutions for two classes of maps:
the conjugates to both the symmetric as well
as asymmetric tent maps. Using these solutions we will derive exact
expressions for important physical quantitites like, for example, the correlation
function. Some specific examples are discussed in detail thereby demonstrating
the deviation of exact and numerically determined properties.

\section{Solutions to the chaotic dynamics of the general conjugates of the symmetric tent map}

\subsection{Trajectories}

Let us begin our investigation by considering the symmetric tent map (STM) \cite{Cau94}.
The $n$-th iterate of the STM $t^{(n)}(x)$ possesses $2^n$ monotonicity intervals with
alternating constant slopes $2^n$ and $-2^n$. The zeros and maximas are at
the positions $\frac{k}{2^n}$  with $k=1,...,2^n$. This makes it possible to 
represent $t^{(n)}(x)$ in the form
\begin{equation}t^{(n)}(x)=\left\{
\begin{array}{llll}
&2^nx-2(k-1) &;\; x\in [\frac{2(k-1)}{2^n},\frac{2k-1}{2^n}] & \\
&-2^nx+2k & ;\;x\in [\frac{2k-1}{2^n},\frac{2k}{2^n}] & \label{tentna}\\
\end{array}\right\}
\end{equation}
The above formula can be recast into the following very simple expression 
\begin{equation}
t^{(n)}(x)=1-\left|(2^n\:x\;{ {\mbox{mod}}}\;2)-1\right|\label{tentn}
\end {equation}
where ${\mbox{mod}}$ is the modulo operation. We consider now the family of maps
$g(x)=u \circ t \circ u^{-1}(x)$
which are obtained from the STM by conjugation with an invertible and
differentiable function $u(x)$ which maps the unit interval onto itself and
obeys $u(0)=0,~u(1)=1$ \cite{Gyo84}. The $n$-th iterates of the maps $\left\{g(x)\right\}$
are then given by
\begin{eqnarray}
g^{(n)}(x)=(u \circ t \circ u^{-1})\circ (u \circ t\circ u^{-1})\circ
\ldots \circ ( u \circ t\circ u^{-1})(x) =u \circ t^{(n)} \circ u^{-1}(x)
\end{eqnarray}
and therefore $g^{(n)}(x)$ takes on the following appearance
\begin{equation}
g^{(n)}(x)=u\left(1-\left|\left(
\left(2^n\:u^{-1}(x)\right)\;{{\mbox{mod}}}\;2\right)-1\right|\right)
\end{equation}
which represents a closed form analytical solution to the dynamics of the
maps conjugated to the STM. The general conjugation yields a variety of 
symmetric as well as nonsymmetric maps. Imposing the restrictive condition
$u(x)=1-u(1-x)$ for the conjugating functions $u(x)$  
we obtain the special class of so-called doubly symmetric maps for which both the invariant
density as well as the map is symmetric. In particular if we use the specific conjugation
$u_l(x)=\sin^2\left(\frac{\pi x}{2}\right)$ and correspondingly 
$u_l^{-1}(x)=\frac{1}{\pi}\arccos(1-2x)$ we arrive at the logistic map $l(x)=4x(1-x)$ and
after a little algebra at the Pincherl$\acute{e}$ relation 
$l^{(n)}(x)=\frac{1}{2}\left\{1-\cos\left[2^n\arccos(1-2x)\right]\right\}$
for the $n$-th iterate of the logistic map. 

Eq.(4) nicely demonstrates the exponential sensitivity with respect to the initial conditions
as well as the infinite process of stretching and folding. The stretching
process takes place through the multiplication with an exponential factor 
($2^n$) and the folding through the ${\mbox{mod}}$-function which cuts with increasing
iterations an increasing number of digits making a more detailed specification
of the initial conditions necessary in order to describe the actual motion.

\subsection{Exact Properties and Examples}

Since the Lyapunov exponent is invariant with respect to conjugation \cite{Cau94,Gyo84b}
all members of the above-discussed class of maps possess the same Lyapunov exponent
$\lambda = \ln 2$. In particular, since $\mu(x)=x$ is the invariant measure of the
STM, the measure of the conjugated maps is $\mu(x)=u^{-1}(x)$ and, therefore, varies
widely with changing conjugating function. Next let us derive analytical expressions
for another important physical quantity namely the correlation function which
is defined by $C(n)=\hat{C}(n)-\bar{x}^2$ with 
$\hat{C}(n)=\int\limits_0^1g^{(n)}(x)x\,d\mu(x)$ and the mean value
$\bar{x}=\int\limits_0^1x\,d\mu(x)$
Let us denote the positions of the zeros and maxima of $g^{(n)}(x)$
by $\left\{x_{2k}\right\}$ and $\left\{x_{2k-1}\right\}$, respectively. 
Using the fact that $g^{(n)}(x)$ is conjugate to some 'original' map
$h^{(n)}(x)$ as well as the relation $u(y_i)=x_i$ the correlation function can be decomposed
as follows
\begin{equation}
\hat{C}(n)=\sum\limits_{k=1}^{2^{n-1}} \hat{C}_k(n)~~~{\mbox{with}}~~~
\hat{C}_k(n)=\int\limits_{y_{2k}}^{y_{2k+1}}u(x)u(h^{(n)}(x))\,dx+
\int\limits_{y_{2k+1}}^{y_{2(k+1)}}u(x)u(h^{(n)}(x))\,dx~~~{n>0}
\end{equation}
In order to provide exact correlation functions for some specific classes of maps
we now specialize to the case that $h(x)$ is given by the STM.
Using eq.(1) we can derive the following simple structure for the terms
of the sum of the correlation function
\begin{eqnarray}  
\hat{C}_k(n)=\int\limits_{\frac{k-1}{2^{n-1}}}^{\frac{2k-1}{2^{n}}}u(x)u(2^nx-2(k-1))\,dx+
\int\limits_{\frac{2k-1}{2^{n}}}^{\frac{k}{2^{n-1}}}u(x)u(-2^nx+2k)\,dx
\end{eqnarray}
In general the terms $\hat{C}_k(n)$ can be evaluated analytically as we shall show in the
following by means of a few examples. Let us first choose the conjugation $u(x)=x^{\frac{1}
{1+\beta}}$ which results in the invariant measure $\mu(x)=x^{1+\beta}$ and the
corresponding normalized power law density $\rho(x) =(\beta+1)x^\beta,\;\;\;\beta>-1$.
Due to their scaling properties power law densities are of particular interest
for physical systems with critical or self-similar behaviour.
Using eqs.(5,6) we arrive after some algebra at the following closed form analytical
expressions for the corresponding correlation function
\begin{eqnarray}C(n)&=&
        \left(\frac{\beta+1}{\beta+3}\right)\left(\frac{1}{2^n}\right)^{\frac{\beta+2}{\beta+1}}+
        2^{\frac{2-n(\beta+2)}{\beta+1}}+
        {}2^{(1-n)\frac{\beta+2}{\beta+1}}\sum\limits_{k=2}^{2^{n-1}}
	\left\{\frac{1}{2}(k-1)^{\frac{1}{1+\beta}}
        B\left(1,\frac{\beta+2}{\beta+1}\right)\right.\;\nonumber\\
	&&\left.{}_2F_1\left( \frac{-1}{1+\beta},\frac{2+\beta}{1+\beta}
        ;\frac{3+2\beta}{1+\beta};\frac{-1}{2(k-1)}\right)+
        2^{\frac{1}{1+\beta}}k^{\frac{\beta+3}{\beta+1}}
        B\left(\frac{2+\beta}{1+\beta},\frac{2+\beta}{1+\beta},\frac{1}{2k}\right)\right\}
        -\left(\frac{\beta+1}{\beta+2}\right)^2
\end{eqnarray}
where $B(a,b)$ and $B(a,b,x)$ denote the complete and incomplete unnormalized
Beta function, respectively, and ${}_2F_1(a,b,c,x)$ is the hypergeometric
function. For the particular case $\beta=0$ the above
expression reduces to the $\delta$-correlation $C(n)\propto\delta_{n0}$ as expected.
Figure 1a shows the absolute value of the above
correlation function for different values of the power $\beta$.
The short time as well as long time behaviour changes somewhat with
changing parameter $\beta$. The asymptotic behaviour $(n\rightarrow\infty)$
of $C(n)$ is an exponential decay.
The decay constant $\tau$ can be determined using the Euler-MacLaurin
sum formula \cite{Ben78} for the asymptotic expansion of eq.(7).
It turns out that for $\beta > 0$ $\tau = \frac{\beta+2}{\beta+1}
\ln 2$ while for $-1 < \beta < 0$ we have $\tau= 2 \ln 2$
independent of $\beta$. 
Having obtained the exact correlation function for the above class of maps
with power law density we are in the position 
to compare these results with those of numerical trajectory simulations. 
Figure 1b shows the comparison of the exact correlation function for
$\beta = 5.0$ (full circles) with the results of numerical trajectory
calculations using $10^3$ (squares) $10^4$ (diamonds) and $10^5$ (triangles) points.
Obviously there is a strong inherent deviation and in particular it 
can be observed that an enhancement of the number of points of the 
numerically calculated trajectory does not yield an improvement in the
statistical accuracy of the corresponding correlation functions. This is 
due to the fact that a naive numerical simulation looses for the above
dynamical systems very rapidly the original trajectory and gets due to 
numerical inaccuracies trapped on a certain 'orbit'.

Finally let us provide one more example out a large number accessible
by the above given analytical formulae. We choose as a conjugating function
$u(x)=\sin(\frac{\pi}{2} x)$ which results in the measure 
$\mu(x)=\frac{2}{\pi} \arcsin(x)$ and the invariant density $\rho(x) =\frac{2}{\pi}
\frac{1}{\sqrt{1-x^2}}$. A straightforward but tedious calculation yields the 
terms $C_k(n)$ which can be summed up to the final beautiful result of the correlation function
\begin{equation}
C(n)=\frac{2^{n+1}}{\pi(2^{2n}-1)}\cot\left(\frac{\pi}{2^{n+1}}\right)-\left(\frac{2}{\pi}\right)^2
\end{equation}
which shows with respect to its asymptotic behaviour
($n\rightarrow\infty$) an exponential decay with a decay constant $2~\ln2$.

\section{Solutions to the chaotic dynamics of the conjugates of the asymmetric tent map}

Solving the problem of the chaotic and ergodic dynamics generated by the general class
of maps conjugated to the asymmetric tent maps (ATM) \cite{Cau94} is a much more intricate
task than the corresponding solution of the dynamics of the maps conjugate to the
STM. In order to develop the necessary concepts and techniques we therefore proceed
in several steps.

\subsection{The Concept of Mode Superposition}

The knowledge of the position of the minima/maxima, i.e. the length of the monotonicity
intervals, of the $n$-th iterate of the ATM is central to the derivation of an analytical
expression for the $n$-th iterate of the ATM. In the present subsection we therefore provide
a method which allows to determine the length $L_n(i)$ of the $i$-th monotonicity interval
(we count the intervals starting from the origin)
by a decomposition technique of the corresponding frequency distribution
$r_n(i)$ (see below) into independent modes.

The ATM is unimodal with constant slopes $1/p$ and $-1/q$ $(q=(1-p))$ on the
two monotonicity intervals, respectively.
First of all we remark that each iteration process $n\rightarrow n+1$ divides a monotonicity
interval of the $n$-th iterate into two monotonicity intervals of the $(n+1)-th$ iterate which
possesses positive and negative slopes in these intervals, respectively.
The ratios of their lengths are given by $p:q$ or $q:p$ depending on whether the $n$-th iterate
of the ATM on the original monotonicity interval possesses a positive
or negative slope, respectively. In order to specify the length of the $i$-th
monotonicity interval we do not need such detailed information like the
symbolic sequence \cite{Cau94} but only the frequency of the occurence of
the factors $p$ and $q$ during the iteration process. The number of different
lengths for the monotonicity intervals is therefore much smaller than the
number of monotonicity intervals themselves. 
The length of the $i$-th monotonicity interval of the $n$-th iterate
of the ATM is given by
\begin{equation}
L_n(i)=p^{n-r_n(i)}q^{r_n(i)}
\end{equation}
where $r_n(i)$ is the frequency of the occurence of the factor $q$ during the 
branching process into the final $i$-th monotonicity interval of the $n$-th iterate.
Using the above-described properties of the branching of the monotonicity intervals
it can be shown that $r_n(i)$ obeys the following recursion formula
\begin{eqnarray}
r_1(1)&=&0 \nonumber\\
r_{n+1}(i)&=&\left \{ \begin{array}{l@{\quad;\quad}c}
r_{n}(i) & 1\le i\le 2^n \\ r_n(2^{n+1}-(i-1))+1 & 2^n+1 \le i \le 2^{n+1}
\end{array}
\right.
\end{eqnarray}
$r_{n+1}(i)$ can therefore be obtained from $r_n(i)$ through a reflection $\sigma_n$
of the function $r_n(i)~(i=1,...,2^n)$ with respect to the vertical axis located at
$2^n+\frac{1}{2}$ and subsequent addition $\tau_n$ of $1$ to the attached part
of $r_{n+1}(i)$ which yields in total
$r_{n+1}(i)=(\tau_n\circ \sigma_n)r_n(i)$. This process is illustrated in Figure 2
for the three functions $r_1(i),r_2(i),r_3(i)$.  

In the following we derive a decomposition of the functions $r_n(i)$ into different
modes, i.e. $r_n(i)$ can then be described by a superposition of these modes.
Let us introduce $n-1$ modes $M_{n,m}(i) \mbox{,} m=2 \ldots n$ and the
additional mode $N_n$. Each mode $M_{n,m}(i)$ is defined on the
finite support $i=1,...,2^n$ and oscillates as a function of $i$ with the
period $T_m=2^m$. In addition it possesses a phase $\phi(M_{n,m})=2^{m-2}$
which describes the shifting of the oscillations on the $i$-axis.
The mode ${N}_n$ is characterized by the period $2^{n}$ and the 
phase $\phi({N}_n)=2^{n-1}$. The equally weighted superposition of these modes
yields the quantity $r_n(i)$, i.e. we have
\begin{equation}
r_n(i)=\sum\limits_{m=2}^n M_{n,m}(i) + {N}_n(i)
\end{equation}
Figure 3 illustrates the three modes $M_{3,2}(i)$, $M_{3,3}(i)$ and ${N}_3(i)$
which are necessary in order to build up $r_3(i)$. In particular we now briefly verify that the
above decomposition obeys the recursion formula given in eq.(10).
The action of the reflection $\sigma_n$ onto the modes $M_{n,m}(i)$ is given by
\begin{equation}
M_{n,m}\stackrel{\sigma_n}{\longrightarrow}M_{n+1,m}~~~~~~~~~~
{N}_n \stackrel{\sigma_n}{\longrightarrow}M_{n+1,n+1}
\end{equation}
The application of the subsequent addition $\tau_n$ to the reflected modes corresponds
to the inclusion of the mode ${N}_{n+1}$. In total we therefore arrive at
the desired relation
\begin{equation}
r_n(i)=\sum\limits_{m=2}^n M_{n,m}(i) + {N}_n(i)\stackrel{\tau_n\circ\sigma_n}{\longrightarrow}
r_{n+1}(i)=\sum\limits_{m=2}^{n+1} M_{n+1,m}(i) + {N}_{n+1}(i)
\end{equation}
In the following subsection we provide construction principles and closed analytical
formulae for the modes $M_{n,m}(i)$ and ${N}_n$.

\subsection{Construction of the Modes}

Each mode function is defined on $2^n$ natural numbers and their
continuation and representation on the real axis is therefore not unique.
In the present subsection we provide two different representations for the mode
functions $M_{n,m}(i)$ and ${N}_n(i)$. The first one is characterized by
the application of the step function to trigonometric functions and the
second one uses polynomials and their periodic continuation in order to describe
the modes.

Looking at Figure 3 suggests a representation of the modes by the action
of the step function ($\Theta(x)=\left \{ 1~:~x\ge0~;~0~:~x<0 \right\}$)
onto an oscillating function. Choosing the $\sin$ function for the oscillating
part we can adapt the periodicity and phase in order to obtain the modes
$M_{n,m}(i)$ and ${N}_n(i)$, respectively. They take on the 
following appearance
\begin{equation}
M_{n,m}(i)=\Theta\left[\sin\left(\pi\frac{2i-(2^{m-1}+1)}{2^m}\right)\right]
~~~~~~~~{N}_n(i)=\Theta\left[\sin\left(\pi\frac{2i-(2^n+1)}{2^n}\right)\right]
\end{equation}
For the function $r_n(i)$ we therefore arrive at the expression
\begin{equation}
r_n(i)=\sum\limits_{m=2}^n \Theta\left[\sin\left(\pi\frac{2i-(2^{m-1}+1)}{2^m}\right)\right]+
\Theta\left[\sin\left(\pi\frac{2i-(2^n+1)}{2^n}\right)\right]
\end{equation}
Introducing the above formula for $r_n(i)$ in eq.(9) we have an analytical
expression for the length of the $i$-th monotonicity interval.
The only feature which could be seen as a disadvantage of the above representation
is its discontinuity through the introduction of the step function. 
It is therefore desirable to gain a second representation of the mode functions
which should have the property of being smooth
with respect to the continuation onto the entire real axis.
We herefore use polynomials whose coefficients will be adapted correspondingly.
The introduction of a $sin$ function yields
then a periodic continuation of the polynomials and allows to adapt
to the functional form of the modes via a corresponding scaling operation and phase shift.

The first step in the derivation of a second, smooth, representation of the modes
$M_{n,m}(i)$ and ${N}_n(i)$ is the construction of a polynomial 
with suitable properties. To achieve this, our starting-point is a polynomial $p_m(x)$
which possesses at $2^{m-1}$ points $x_i^{+}$ the value $+1$ and at
other $2^{m-1}$ points $x_i^{-}$ the value $-1$. 
\begin{equation}p_m(x_1^+)=p_m(x_2^+)= \ldots =p_m(x_{2^{m-1}}^+)=1~~~~~~~
p_m(x_1^-)=p_m(x_2^-)= \ldots =p_m(x_{2^{m-1}}^-)=-1\end{equation}
For the time being the domain of definition of the above introduced polynomial
includes $2^m$ integer values and possesses $2^{m-1}$ zeros.
Such a polynomial can be written as a sum of $2^m$ terms, each of the terms
thereby acquires the value $+1$ or $-1$ at exactly one position given by
$x_i^{+}$ or $x_i^{-}$, respectively whereas at all other positions 
$x_j^+$,$x_j^-$, $i \not= j$ it vanishes. These properties are guaranteed
if each term consists of a normalized product of linear factors and
we therefore arrive at the following expression for the polynomial $p_m(x)$
\begin{equation}
p_m(x) = \sum\limits_{i=1}^{2^{m-1}}\frac{(x-x_i^-)}{(x_i^+-x_i^-)}\prod
\limits_{j=1 \atop j \not=i}^{2^{m-1}}\frac{(x-x_j^+)(x-x_j^-)}{(x_i^+-x_j^+)(x_i^+-x_j^-)}
-\sum\limits_{i=1}^{2^{m-1}}\frac{(x-x_i^+)}{(x_i^--x_i^+)}
\prod\limits_{j=1 \atop j \not=i}^{2^{m-1}}\frac{(x-x_j^+)(x-x_j^-)}{(x_i^--x_j^+)(x_i^--x_j^-)}
\end{equation}
The points of support $x_i^+$, $x_i^-$ ought to be equidistant 
in the unit interval $[0,1]$ and should be arranged symmetrically with respect
to $0$, i.e. we have
\begin{equation}
x_i^-=-x_i^+=x_i~~~~~~~~~x_i=\frac{2i-1}{2^m}
\end{equation}
After a little algebra this leads to an essential simplification of the polynomials (17)
\begin{equation}
p_m(x)=\sum\limits_{i=1}^{2^{m-1}} \frac{x}{x_i}\prod
\limits_{j=1 \atop j \not=i}^{2^{m-1}}\frac{x^2-x_j^2}{x_i^2-x_j^2}
\end{equation}
As a next step we perform a periodic continuation of the polynomial $p_m(x)$
by substituting
\begin{equation}
x\longrightarrow\sin(\pi x)~~~~~~~~x_i\longrightarrow\sin(\pi x_i)
\end{equation}
In order to describe the modes with the constructed polynomials we have
to adapt the frequency as well as phase of the oscillating periodic functions defined
by eqs.(19,20). In addition a shift is performed in order to make the points of support
equal to integers on which the modes are defined. Finally we arrive at the following
expressions for the individual modes
\begin{equation}
M_{n \mbox{,} m}(x)=\frac{1}{2} \left\{1+\sum\limits_{i=1}^{2^{m-1}} \frac{\sin\left(\pi
\frac{2x-(2^{m-1}+1)}{2^m}\right)}{\sin\left(\pi \frac{2i-1}{2^m}\right)}\prod
\limits_{j=1 \atop j \not=i}^{2^{m-1}}
\frac{\sin^2\left(\pi \frac{2x-\left(2^{m-1}+1\right)}{2^m}\right)-
\sin^2\left(\pi \frac{2j-1}{2^m}\right)}
{\sin^2\left(\pi \frac{2i-1}{2^m}\right)-\sin^2\left(\pi \frac{2j-1}{2^m} \right )
}\right \}
\end{equation}
\begin{equation}
{N}_n(x)=\frac{1}{2} \left\{1+\sum\limits_{i=1}^{2^{n-1}} \frac{\sin\left(\pi
\frac{2x-(2^n+1)}{2^n}\right)}{\sin\left(\pi \frac{2i-1}{2^n}\right)}\prod
\limits_{j=1 \atop j \not=i}^{2^{n-1}}
\frac{\sin^2\left(\pi \frac{2x-\left(2^n+1\right)}{2^n}\right)-\sin^2\left(\pi \frac{2j-1}{2^n}\right)}
{\sin^2\left(\pi \frac{2i-1}{2^n}\right)-\sin^2\left(\pi \frac{2j-1}{2^n} \right )
}\right \}
\end{equation}
and for the frequency distribution $r_n(i)$ of the $i$-th monotonicity interval

\begin{eqnarray}
r_n(i)&=&\sum\limits_{m=2}^n  \frac{1}{2} \left\{1+\sum\limits_{i=1}^{2^{m-1}}
\frac{\sin\left(\pi \frac{2x-(2^{m-1}+1)}{2^m}\right)}{\sin\left(\pi 
\frac{2i-1}{2^m}\right)}\prod \limits_{j=1 \atop j \not=i}^{2^{m-1}}
\frac{\sin^2\left(\pi \frac{2x-\left(2^{m-1}+1\right)}{2^m}\right)-
\sin^2\left(\pi \frac{2j-1}{2^m}\right)} {\sin^2\left(\pi \frac{2i-1}{2^m}\right)
-\sin^2\left(\pi \frac{2j-1}{2^m} \right ) }\right \}\nonumber \\
&&+\frac{1}{2} \left\{1+\sum\limits_{i=1}^{2^{n-1}} \frac{\sin\left(\pi
\frac{2x-(2^n+1)}{2^n}\right)}{\sin\left(\pi \frac{2i-1}{2^n}\right)}\prod
\limits_{j=1 \atop j \not=i}^{2^{n-1}}
\frac{\sin^2\left(\pi \frac{2x-\left(2^n+1\right)}{2^n}\right)
-\sin^2\left(\pi \frac{2j-1}{2^n}\right)}
{\sin^2\left(\pi \frac{2i-1}{2^n}\right)-\sin^2\left(\pi \frac{2j-1}{2^n}
\right ) }\right \} 
\end{eqnarray}
This concludes our construction of the modes.

\subsection{Analytical Representation of the Chaotic Dynamics}

The knowledge of the frequency distribution $r_n(i)$ is a key ingredient for the
derivation of analytical solutions to the chaotic motion of the conjugates of the
ATM. Arbitrarily high iterates are now accessible in closed analytical form via
the following procedure. 

First of all we give some important quantities related to the $n$-th iterated map:

\begin{itemize}
\item The length of the $i$-th monotonicity intervals is given by
        \begin{equation}L_n(i)=p^{n-r_n(i)}q^{r_n(i)}\end{equation}
\item The slope in the $i$-th monotonicity interval is
\begin{equation}
S_n(i)=p^{r_n(i)-n}q^{-r_n(i)} (-1)^{(i-1)}
\end{equation}
\item The left and right boundary $G^L_n(i)$ and $G^R_n(i)$ of the $i$-th monotonicity
interval can be obtained by summation of the lengths of the corresponding intervals:
\begin{equation}
G^L_n(i)=\sum\limits_{j=1}^{i-1} L_n(j)=\sum\limits_{j=1}^{i-1} p^{n-r_n(j)}q^{r_n(j)}
~~~~~~~G^R_n(i)=\sum\limits_{j=1}^i L_n(j)=\sum\limits_{j=1}^i p^{n-r_n(j)}q^{r_n(j)}
\end{equation}
\item Definition of the index-function:
Let us define a function $I_n:~[0,1] \rightarrow \left\{1,...,2^{n}\right\}$ whose value $I_n(x)$
tells us the number of the monotonicity interval in which $x$ is contained.
The index function can be described with the help of the step function $\Theta(x)$
in the following way
\begin{equation}
I_n(x)=\sum\limits_{i=1}^{2^n}\Theta \left [x-G^L_n(i)\right ]=
\sum\limits_{i=1}^{2^n}\Theta \left [x-
\sum\limits_{j=1}^{i-1} p^{n-r_n(j)}q^{r_n(j)}\right ]
\end{equation}
\end{itemize}
We are now in the position of formulating the $n$-th iterate $t_p^{(n)}(x)$ of the
ATM in a closed analytical expression
\begin{equation}
t_p^{(n)}(x)=\left \{x-G^L_n[I_n(x)]\right \} S_n[I_n(x)]+\frac{1}{2}\left(1+(-1)^{I_n(x)}
\right)
\end{equation}
Inserting eqs.(26,27) into the above expression yields
\begin{equation}
t_p^{(n)}(x)=\left [x-\sum\limits_{j=1}^{I_n(x)-1} p^{n-r_n(j)}q^{r_n(j)}\right ]
p^{r_n[I_n(x)]-n}q^{-r_n[I_n(x)]} (-1)^{I_n(x)-1}
+\frac{1}{2}\left(1+(-1)^{I_n(x)}\right)
\end{equation}
Similar to the case of the STM we obtain the $n$-th iterate $g_{t_p}^{(n)}(x)$
of the maps conjugated to the ATM by the following decomposition
\begin{equation}
g_{t_p}^{(n)}(x)=u\circ t_p \circ u^{-1}\circ u\circ t_p \ldots \circ t_p \circ u^{-1}(x)
=u\circ t_p ^{(n)}\circ u^{-1}(x)
\end{equation}
which gives together with eq.(29) our final result
\begin{equation}
g_{t_p}^{(n)}(x)=u\left\{\left [u^{-1}(x)-\sum\limits_{j=1}^{I_n(u^{-1}(x))-1}
L_n(j)\right ]
S_n\left(I_n(u^{-1}(x))\right) +\frac{1}{2}\left(1+(-1)^{I_n(u^{-1}(x))}\right)\right\}
\end{equation}
Eq.(31) can be used to calculate arbitrarily high iterates
of the conjugates of the ATM. As an example we show in figure 4 the $8$-th iterate
of the conjugate $u(x)=\sin^2\left(\frac{\pi x}{2}\right)$ of the
ATM for $p=0.9$ which gives an idea of the scaling structures contained 
in the above analytical formula. 

The invariant measure of the conjugates of the ATM is given by $\mu(x)=u^{-1}(x)$
and the Lyapunov exponent is known to be $\lambda = -p \ln(p) -(1-p) \ln(1-p)$
\cite{Mor81}. The above eq.(31) can be used to calculate the correlation
functions of the conjugates of the ATM as well as any other physical property.

\section{Conclusions}

We have presented analytical solutions to the chaotic and ergodic dynamics
of two classes of noninvertible single humped maps: the conjugates of the
symmetric as well as asymmetric tent map. With the exception of the
iterates of the logistic map this is, to our knowledge, the first 
time that closed analytical representations of chaotic trajectories  
are derived. This enables us to calculate exact physical quantities,
like, for example, the time correlation function, even in those cases
for which numerical trajectory calculations fail to predict the correct long time
behaviour. The complex stretching and folding process of the iteration of the
dynamical system is clearly revealed within our analytical approach
through the combination of conjugating functions, scale transformations
and cut-off operations.

The key to the construction of the analytical solutions is the complex
superposition of modes connected with the monotonicity intervals of the
$n$-th iterate of the dynamical system. We conjecture that this holds
also for the general case of arbitrary single humped maps and might be
a hint towards the construction of their solutions.
The boundaries of the monotonicity intervals are then given
by the preimages of the maximum of the dynamical law.

\vspace*{2.0cm}

\begin{center}
{\bf{{FIGURE CAPTIONS}}}
\end{center}

\noindent
{\bf{Figure~1}}:~(a) The logarithm of the absolute value of the correlation function for
the conjugation $u(x)=x^{\frac{1}{1+\beta}}$ of the STM for different
values of the parameter $\beta$. The values $\beta = -0.95,-0.5,+1.0,+5.0,+50.0$
correspond to the curves with the full circles, circles, plus, full triangles and
full boxes, respectively.
(b) The exact correlation function for $\beta = 5.0$ (solid line) together with the correlation
functions resulting from numerical trajectory calculations (broken lines) with 
$10^3$ (squares), $10^4$ (diamonds) and $10^5$ (triangles) points. 

\vspace*{0.5cm}

\noindent
{\bf{Figure~2}}:~The frequency distributions $r_1(i),r_2(i)$ and $r_3(i)$ of the ATM.

\vspace*{0.5cm}

\noindent
{\bf{Figure~3}}:~The modes $M_{3,2}(i),M_{3,3}(i)$ and $N_3(i)$ of the ATM.

\vspace*{0.5cm}

\noindent
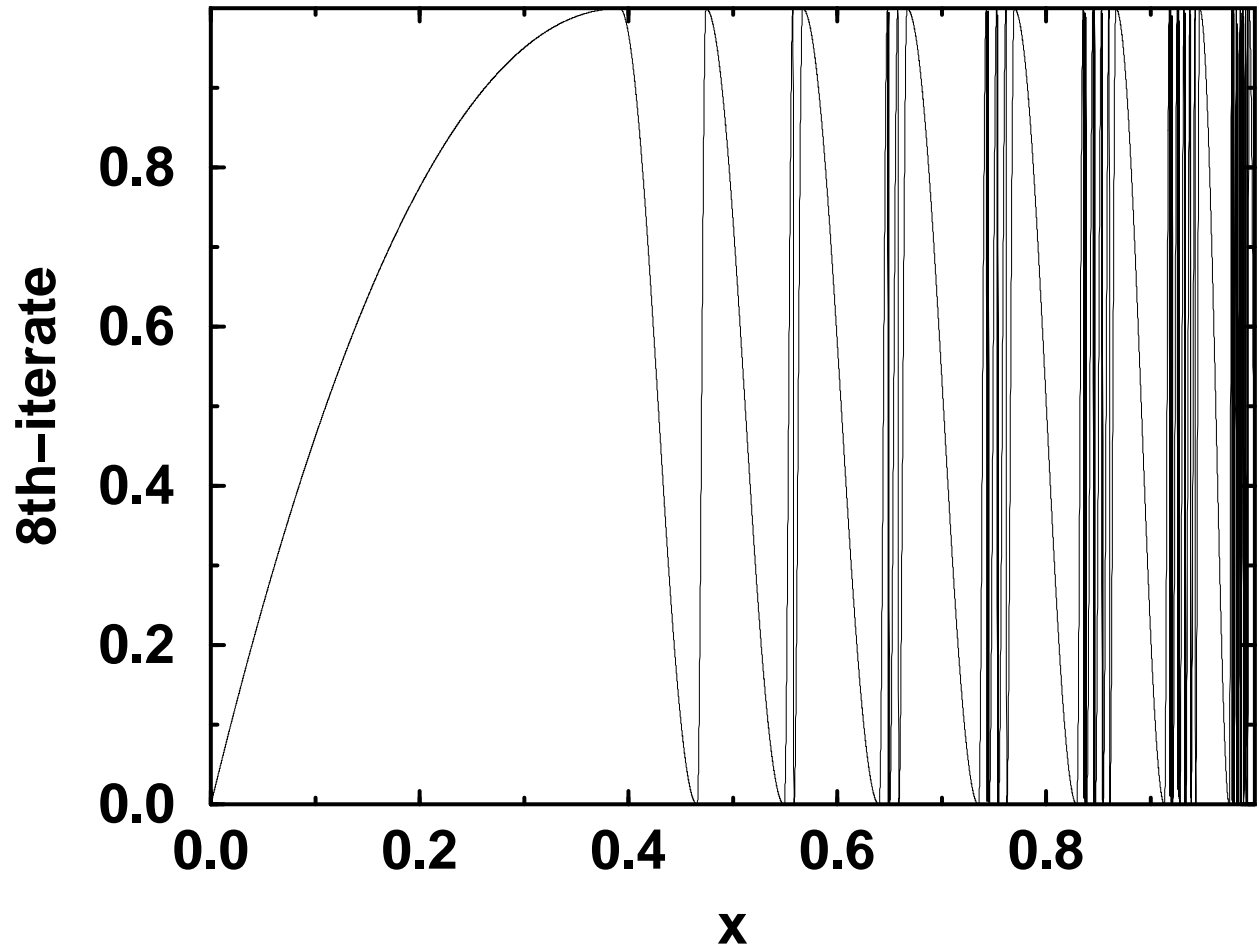
{\bf{Figure~4}}:~The $8$-th iterate of the conjugate
$u(x)=\sin^2\left(\frac{\pi x}{2}\right)$ of the ATM for $p=0.9$ calculated through
eq.(31).


{}

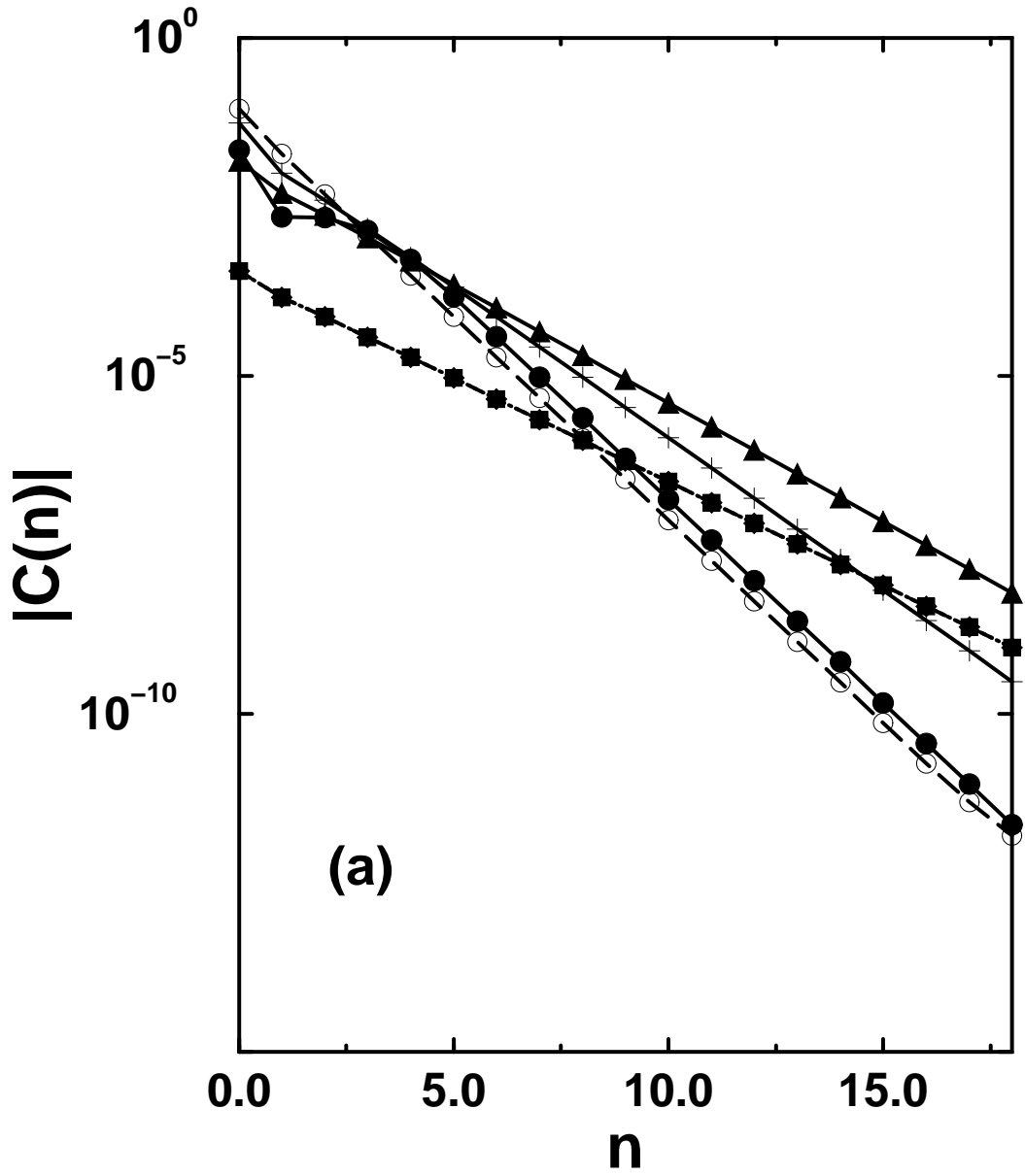
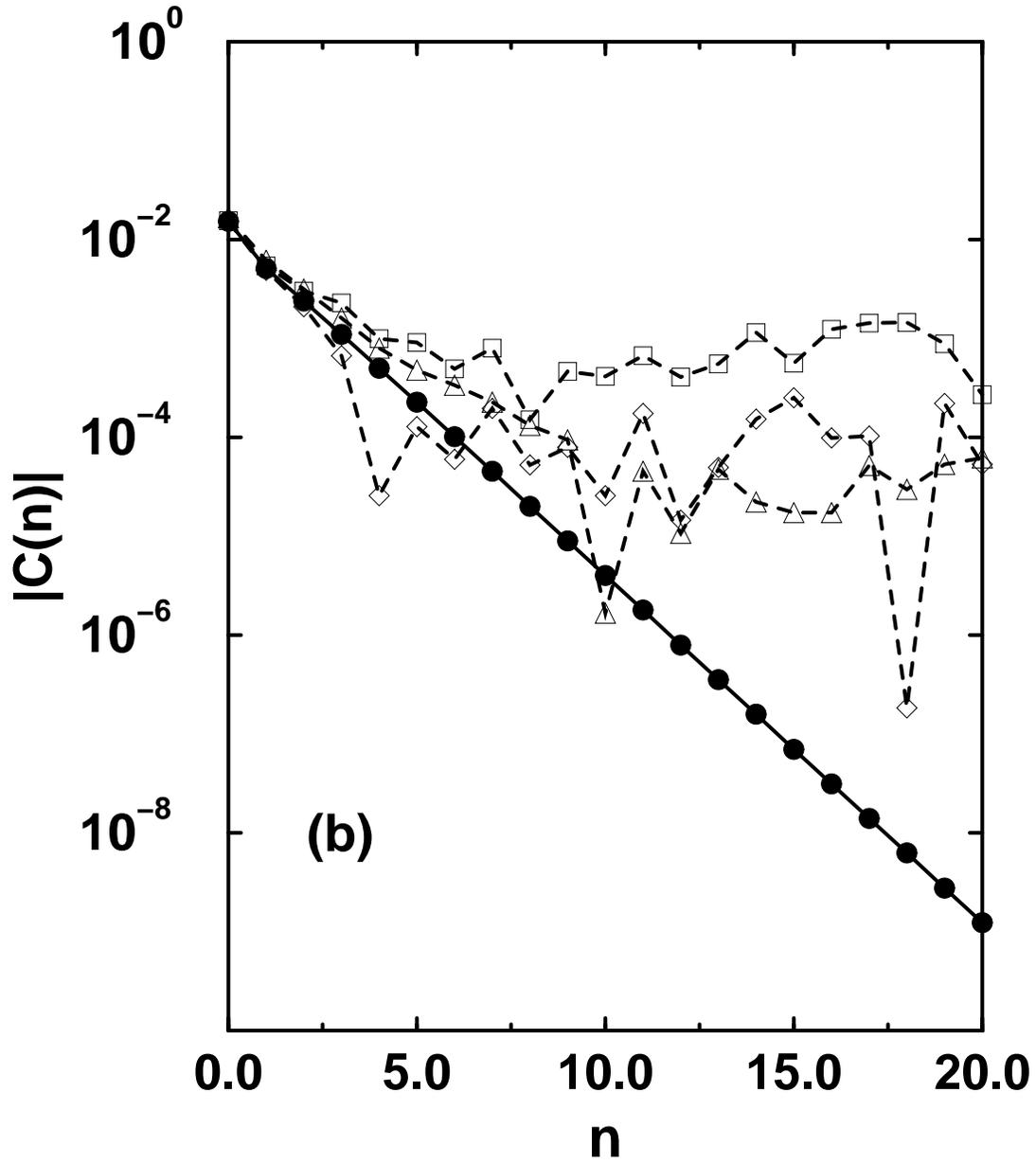
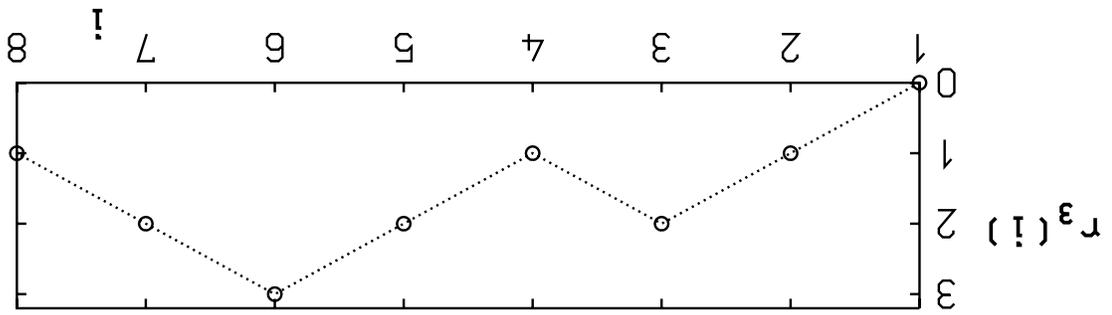
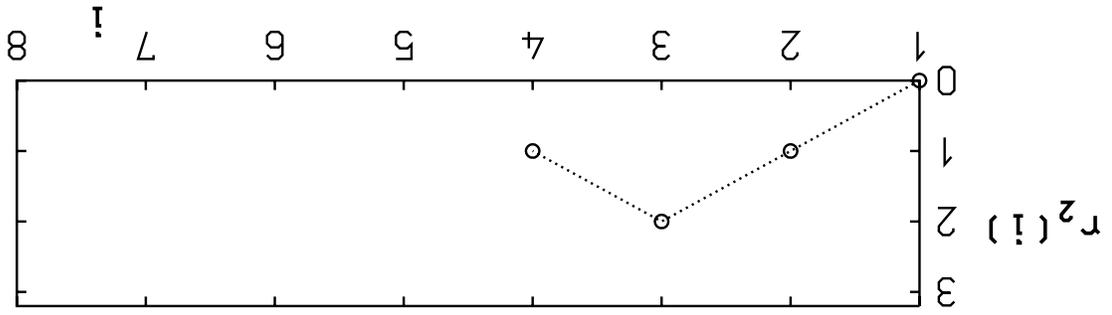
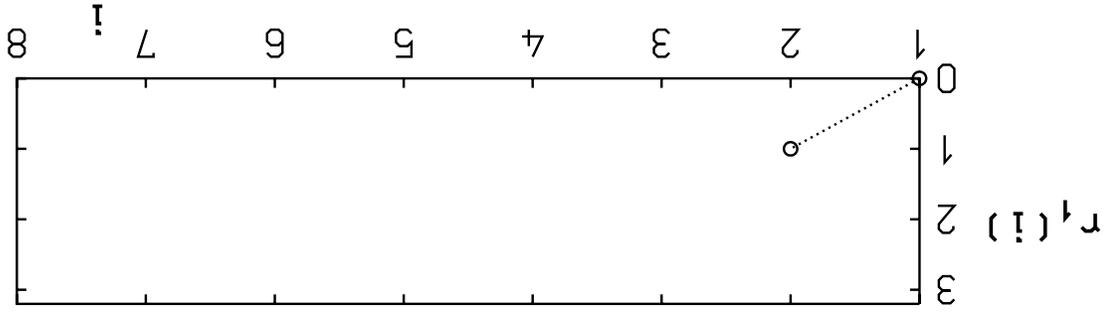
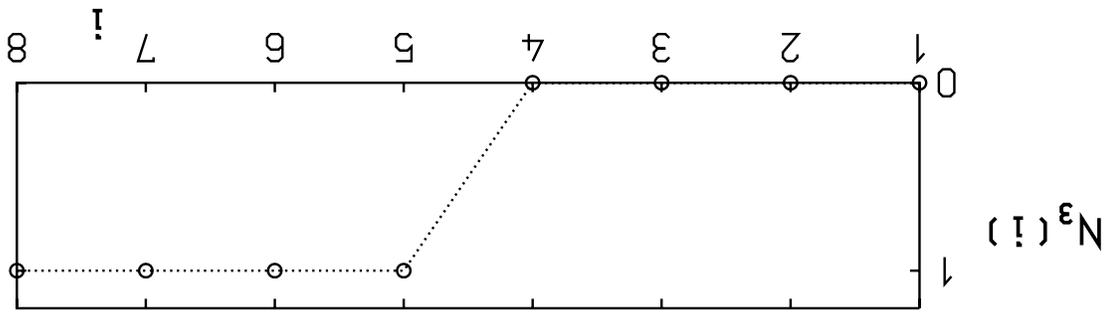
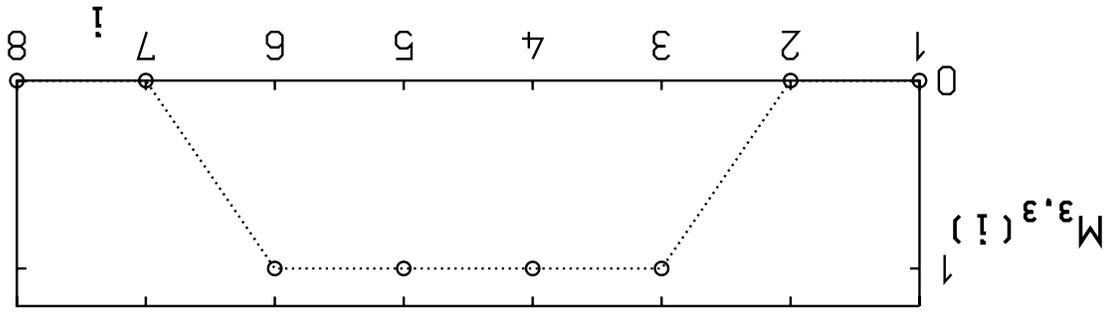
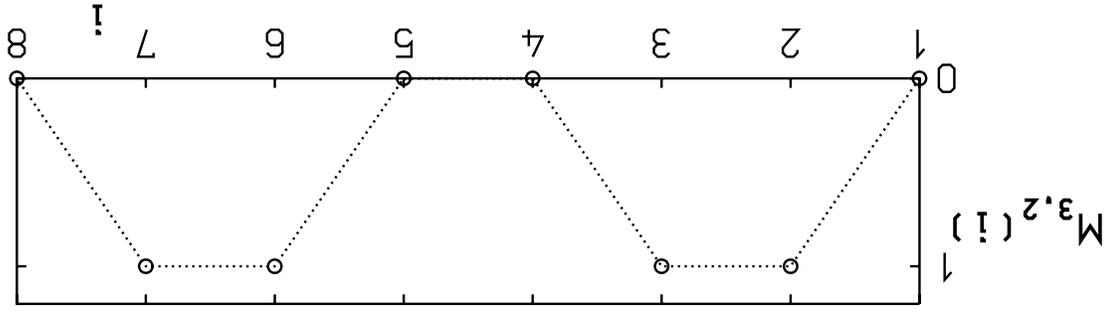
\end{document}